\documentclass[twocolumn]{aastex61}

\usepackage{rotating,natbib,epsfig}
\submitjournal{Astronomical Journal}

\shorttitle{New Binaries in $\epsilon$ Cha} 

\shortauthors{Brice\~no \& Tokovinin}

\begin{document}

% for float placement:
\renewcommand{\topfraction}{1.0}
\renewcommand{\bottomfraction}{1.0}
\renewcommand{\textfraction}{0.0}

\title{New binaries in the $\epsilon$ Cha  association \footnote{Based on observations obtained  at the Southern Astrophysical Research
(SOAR) telescope. }}
% a  joint project of the  Minist\'{e}rio da
%Ci\^{e}ncia,  Tecnologia, e  Inova\c{c}\~{a}o (MCTI)  da Rep\'{u}blica
%Federativa do Brasil, the  U.S. National Optical Astronomy Observatory
%(NOAO), the  University of  North Carolina at  Chapel Hill  (UNC), and
%Michigan State University (MSU).

\author{C\'esar Brice\~no}
\author{Andrei Tokovinin}
\affiliation{Cerro Tololo Inter-American Observatory, Casilla 603, La Serena 1700000, Chile}
\email{cbriceno@ctio.noao.edu, atokovinin@ctio.noao.edu}
\correspondingauthor{C\'esar Brice\~no}

\begin{abstract}
We  present Adaptive  Optics-aided  speckle observations  of 47  young
stars  in the  $\epsilon$ Cha  association  made at  the 4-m  Southern
Astrophysical Research Telescope  in the $I$ band.  We  resolved 10 new
binary pairs, 5 previously known binaries and two triple systems, also
previously known.  In  the separation range between 4 and   300 AU, the 30
association  members of  spectral  types  G0 and  later  host 6  binary
companions, leading  to the raw companion  frequency of 0.010$\pm$0.04
per  decade of  separation,  comparable  to the  main sequence
dwarfs in the field. On the other hand, all 5 massive association
members of  spectral types A and  B have companions in  this range. We
discuss the newly resolved and known binaries in our sample.  Observed
motions in the triple system  $\epsilon$ Cha, composed of three similar
B9V  stars, can be  described  by  tentative orbits  with  periods 13  and
$\sim$900 years, and a large mutual inclination. 
\end{abstract} 

%---------------------------------------------------------
\section{Introduction}
\label{sec:intro}

Multiple  star systems are  a common  product of  the process  of star
formation     \citep{duchene_kraus2013}.     Characterizing    stellar
multiplicity in  young stellar  populations is therefore  an important
and necessary step toward  understanding issues like the fragmentation
of primordial cores or massive disks, the early dynamical evolution of
stellar  systems, the  initial mass  function of  single  stars versus
multiples,  and  how stellar  multiplicity  affects  the survival  and
evolution  of  circumstellar  disks,  which bears  ultimately  on  the
frequency and properties of planets orbiting binary and multiple star
systems \citep[e.g., see review by ][]{reipurth_etal2014}.

Over the  past decades  we have learned  that loose  associations like
Taurus and Chamaeleon\,I harbor roughly twice as many low-mass ($M \la
2 \>  M_{\odot}$) pre-main sequence binaries than  compact clusters of
similar  age   like  the  Orion  Nebula  Cluster   (ONC),  which  have
multiplicity       fractions      similar      to       the      field
\citep[e.g.,][]{petr1998,koh2006}.    These    observations   can   be
interpreted with the assumption  of an universally high ($\sim 100$\%)
primordial multiplicity  fraction for  all star forming  regions, with
the subsequent  rapid dynamical disruption of binaries  in young dense
clusters,  which  significantly   lowers  the  multiplicity  fractions
\citep{kroupa1995,kroupa1999,kroupa2011}.   However,  the universality
of a  high primordial multiplicity  fraction, independent of  the star
forming environment,  has lately  been questioned and  remains debated
\citep{king2012,marks2014}.   \cite{parker2014}  find  that  dynamical
processing of  populations composed of  100\% binaries, even  in dense
star  forming regions,  cannot explain  the clear  differences  in the
Galactic field  binary fraction and  mean separation as a  function of
decreasing primary  mass.  In summary, there  is accumulating evidence
that the  primordial binary frequency and  separation distribution are
{\it not} universal,  but may depend on the  star forming environment.
Therefore it  is important to measure multiplicity  properties of many
young clusters and associations. This work focuses on one such group.

Because of  their proximity, the so-called nearby  young moving groups
\citep{zuckerman2004,torres_etal2008},  like the $\eta$  Cha cluster \citep{mamajek1999} 
and $\epsilon$ Cha  association, are  excellent laboratories  to  investigate stellar
multiplicity  over a  wide range  of separations.   These  two stellar
aggregates were first proposed by \cite{frink_etal1998} as a kinematic
group of young  stars in the general direction  of the Chamaeleon dark
clouds.   \cite{mamajek_etal2000} discuss $\epsilon$  Cha as  a sparse
association  in  the context  of  other  nearby  stellar systems,  and
characterize  it  as   a  distinct  group  of  5-15   Myr  old  stars.
\cite{fwg2003} derived an age of  3-5 Myr for $\epsilon$ Cha.  Several
recent studies have concentrated  on building better membership lists.
In  their  re-examination  of  the $\epsilon$  Cha  group  membership,
\cite{Murphy2013} arrive at a final list of 35-41 members, with a mean
distance of $110 \pm 7$ pc and an age $\sim 3-5$ Myr, making it likely
the youngest  of the nearby moving  groups. This is  the most complete
census  of   the  association  available  to   date;  for  comparison,
\replaced{Elliot et al. (2016) listed only 17 members of 
$\epsilon$~Cha}{\citet{Elliott2015} listed only 24 members of $\epsilon$~Cha \citep[17
of  them are found in][]{Murphy2013}. } Note that
attribution  of a  star  to a  particular  association is  complicated
because of the partial overlap between young groups; binarity adds yet
another complication by distorting photometry and proper motions.

Multiplicity  of  young   stars  has  been  extensively  characterized
observationally, mostly by  high angular resolution imaging \citep[see
  the  review in][]{duchene_kraus2013}.   However, $\epsilon$  Cha has
been largely  neglected so far  by these studies.  The  most extensive
data  are provided  by  \citet{koh2001b} who  observed X-ray  selected
young  stars  in the  direction  of  the  dark clouds  in  Chamaeleon,
resolving binaries  with separations  from 0\farcs13 to  6\arcsec.  He
made no distinction  between the more distant Cha~I  and Cha~II groups
and the foreground $\epsilon$~Cha association; 18 objects of his study
overlap with our sample.   K\"ohler found the multiplicity fraction to
be  comparable  to  the  field.  Recently  \citet{Elliott2015}  probed
binarity in  $\epsilon$ Cha by high-resolution imaging  of 10 targets,
detecting three binaries.  One of those, 
\replaced{KOH-93)}{RXJ1220.4-7407 (KOH~93)}, was first resolved by
K\"ohler,  TYC  9245-535-1 is  not featured  in our  input list,
while HD~105923 is independently confirmed by our survey.

Here  we present  a study  of multiplicity  of 47  young stars  in the
direction  of  the $\epsilon$  Cha  stellar  group;  37 of  those  are
confirmed or candidate  members of the association.  This  is the most
extensive search for binaries  and multiple systems in $\epsilon$ Cha,
in the  separation range $\sim 0\farcs04 -  3\arcsec$ (equivalent to
projected separations $\sim  4 - 300$ AU, assuming  a mean distance of
100 pc).  In \S {\ref{sec:obs}} we present the observations, and in \S
\ref{sec:res}  our   results.   Section~\ref{sec:disc}  discusses  the
multiplicity in  $\epsilon$~Cha and concludes the paper.

%---------------------------------------------------------
\section{Observations}
\label{sec:obs}

%-------------------------------------------------------------

\subsection{Sample Selection}

We used for  the target selection the list of  proposed members of the
$\epsilon$  Cha association  from the  Table~1  of \citet{Murphy2013}.
Here in  Table~\ref{targets} we provide the characteristics  of the 47
observed stars. \replaced{The members  confirmed by \citet{Murphy2013} have the
	Cha-NN numbers in the second  column; the candidate members are marked
	as Cha-cand, while  the rejected members are labeled  as Cha-rej.  Two
	candidates, TYC  9414-191-1 and TYC 9420-676-1, can  be rejected based
	on their {\it Gaia} parallaxes, so their status is set accordingly. On
	the other hand, HIP~55746 has {\it Gaia} parallax of 10.74\,mas and is
	most  likely   a  member  of  the  association;   its  rejection  by
	\citet{Murphy2013} could be caused by its close binary companion.  For
	previously  known  multiples  listed  in the  Washington  Double  Star
	Catalog, WDS  \citep{WDS}, the ``discoverer  codes'' are given  in the
	second   column.    The   separations    in   the   last   column   of
	Table~\ref{targets}  indicate   which  objects  have   been  resolved;
	asterisks mark the first-time resolutions.}{The members confirmed by \citet{Murphy2013} 
	have their corresponding Cha-NN numbers in column (2); candidate members are marked
as Cha-cand, while  the rejected members are labeled  as Cha-rej. 
Columns (2) and (3) contain the star position, columns (5) and (6) the
$I_c$ magnitude and Spectral Type respectively, and in columns (7) and (8)
we provide distances derived from the {\it Gaia} DR1 parallaxes \citep{gaia2016a},
and from \cite{Murphy2013}. 
Two candidates, TYC  9414-191-1 and TYC 9420-676-1, can  be rejected based
on their {\it Gaia} parallaxes  \added{\citep{Gaia}}, so their status is
set  accordingly. On  the other  hand,  HIP~55746 has  a {\it  Gaia}
parallax of 10.74\,mas and is most likely a member of the association;
its  rejection by  \citet{Murphy2013}  could have been caused  by its  close
binary  companion.   For  previously  known multiples  listed  in  the
Washington  Double  Star Catalog,  WDS  \citep{WDS}, the  ``discoverer
codes'' are given  in column (2).  The  separations in column (9)
of  Table~\ref{targets}   indicate  which  objects  have  been
resolved;  asterisks  mark  first-time  resolutions. Three
  stars from  the original list, 2MJ11334926-7618399, 2MJ11404967-7459394 and
2MJ12014343-7835472, all fainter than $I_c = 14.1$, were not observed 
and therefore are not included in Table~1.}

The final list contains 37 $\epsilon$~Cha members/candidates among the
47 entries in Table~1.  However, it  should be pointed out that the 10
rejected  stars  are,  for   the  most  part,  young  objects.   Their
attribution to a particular  association or group might be compromised
by  multiplicity.   We  believe  that their  observations  are  useful
regardless of the membership  status.  Monitoring of known close young
binaries     will     establish     their    orbits     and     masses
\citep[e.g. EG~Cha,][]{Tok2016}.

\tabletypesize{\footnotesize}
\tablewidth{0pt}
\startlongtable
\begin{deluxetable*}{llccccccc}
\tablecaption{Targets observed in the $\epsilon$ Cha stellar group \label{targets}}
\tablehead{
		\colhead{NAME} & \colhead{Membership} & \colhead{$\rm RA$(J2000)} & \colhead{$\rm DEC$(J2000)} & \colhead{$I_{\rm c}$} & \colhead{SpT} & \colhead{Distance\tablenotemark{a}} & \colhead{Distance\tablenotemark{b}} & \colhead{Sep.} \\
		\colhead{} & \colhead{Other ID} & \colhead{(h : m : s)}  & \colhead{($^\circ$ : $\arcmin$ : $\arcsec$)} & \colhead{(mag)} &  \colhead{} & \colhead{(pc)} & \colhead{(pc)}  & \colhead{(\arcsec)} 	}
\colnumbers
\startdata
HD 82879              & Cha-rej               & 09:28:21.1 & -78:15:35.0 & 8.95  & F6   & $111\pm 7$  & \ldots & \ldots \\
CP-68 1388            & Cha-22                & 10:57:49.3 & -69:14:00.0 & 9.28  & K1   & $114\pm 6$  & 112    & \ldots \\
VW Cha                & Cha-rej; GHE 35       & 11:08:01.5 & -77:42:29.0 & 11.03 & K8   & \ldots      & \ldots &  0.67  \\
TYC 9414-191-1        & Cha-rej               & 11:16:29.0 & -78:25:20.8 & 9.59  & K5   & $477\pm 157$& 105    & \ldots \\
2MJ11183572-7935548   & Cha-13                & 11:18:35.7 & -79:35:54.8 & 12.22 & M4.5 & \ldots      & 101    & 0.92*  \\
RXJ1123.2-7924        & Cha-14,rej            & 11:22:55.6 & -79:24:43.8 & 11.62 & M1.5 & \ldots      & \ldots & \ldots \\
HIP 55746             & Cha-cand              & 11:25:18.1 & -84:57:16.0 & 7.15  & F5   & $93\pm 6$   & \ldots & 0.06*  \\
RXJ1137.4-7648        & Cha-rej               & 11:37:31.3 & -76:47:59.0 & 12.2  & M2.2 & \ldots      & \ldots & 2.89*  \\
TYC 9238-612-1        & Cha-rej               & 11:41:27.7 & -73:47:03.0 & 9.98  & G5   & $158\pm 6$  & \ldots & 2.28*  \\
2MJ11432669-7804454   & Cha-17                & 11:43:26.7 & -78:04:45.4 & 13.51 & M4.7 & \ldots      & 117    & \ldots \\
RXJ1147.7-7842        & Cha-23                & 11:47:48.1 & -78:41:52.0 & 10.92 & M3.5 & \ldots      & 106    & \ldots \\
RXJ1149.8-7850        & Cha-18; DZ Cha        & 11:49:31.8 & -78:51:01.1 & 11.01 & M0   & \ldots      & 110    & \ldots \\
RXJ1150.4-7704        & Cha19,rej             & 11:50:28.3 & -77:04:38.0 & 10.54 & K4   & \ldots      & \ldots & \ldots \\
RXJ1150.9-7411        & Cha-24; BRR 15        & 11:50:45.2 & -74:11:13.0 & 12.06 & M3.7 & \ldots      & 108    &  0.91  \\ 
2MJ11550485-7919108   & Cha-25                & 11:55:04.9 & -79:19:10.9 & 13.26 & M3   & \ldots      & 115    & \ldots \\
T Cha                 & Cha-26; HIP 58285     & 11:57:13.5 & -79:21:31.5 & 10.28 & K0   & $107\pm 7$  & 108    & \ldots \\
RXJ1158.5-7754B       & Cha-20;               & 11:58:26.8 & -77:54:45.0 & 11.81 & M3   & \ldots      & 119    & \ldots \\
RXJ1158.5-7754A       & Cha-21; HIP 58400     & 11:58:28.2 & -77:54:29.6 & 9.76  & K4   & $90\pm 5$   & 104    &  0.03  \\
HD 104036             & Cha-27; EE Cha        & 11:58:35.2 & -77:49:32.0 & 6.49  & A7   & $108\pm 8$  & 108H   &  0.63* \\
CXOUJ115908.2-781232  & Cha-1,cand            & 11:59:08.0 & -78:12:32.2 & 13.83 & M4.75& \ldots      & 165    & \ldots \\
$\epsilon$ Cha        & Cha-2; HJ 4486        & 11:59:37.5 & -78:13:18.9 & 5.39  & B9   & \ldots      & 111H   & Triple \\
RXJ1159.7-7601        & Cha-28; HIP 58490     & 11:59:42.3 & -76:01:26.2 & 10.18 & K4   & $101\pm 5$  & 107    & \ldots \\
HD 104237A            & Cha-5; GRY 1          & 12:00:05.1 & -78:11:34.6 & 6.31  & A7.75& $104\pm 6$  & 114H   &  1.39  \\
HD 104237D            & Cha-6; FGL 2          & 12:00:08.3 & -78:11:39.6 & 11.62 & M3.5 & \ldots      & 114H   &  4.2   \\
HD 104237E            & Cha-7; FGL 2          & 12:00:09.3 & -78:11:42.5 & 10.28 & K5.5 & \ldots      & 114H   & \ldots \\
2MJ12005517-7820296   & Cha-10                & 12:00:55.2 & -78:20:29.7 & 14.0  & M5.75& \ldots      & 126    & \ldots \\
HD 104467             & Cha-29                & 12:01:39.1 & -78:59:16.9 & 7.81  & G3   & $95\pm 6$   & 102    & \ldots \\
USNOB 120144.7-781926 & Cha-8                 & 12:01:44.4 & -78:19:26.8 & 13.72 & M5   & \ldots      & 100    & \ldots \\
CXOUJ120152.8-781840  & Cha-9                 & 12:01:52.5 & -78:18:41.4 & 13.52 & M4.75& \ldots      & 121    & \ldots \\
RXJ1202.1-7853        & Cha-30                & 12:02:03.8 & -78:53:01.0 & 10.49 & M0   & \ldots      & 110    &  0.05* \\
RXJ1202.8-7718        & Cha-cand              & 12:02:54.6 & -77:18:38.2 & 11.9  & M3.5 & \ldots      & 128    & \ldots \\
RXJ1204.6-7731        & Cha-31                & 12:04:36.1 & -77:31:34.6 & 11.25 & M3   & \ldots      & 112    & \ldots \\
TYC 9420-676-1        & Cha-rej               & 12:04:57.4 & -79:32:04.7 & 9.73  & F0   & $248\pm 47$ & \ldots &  0.65* \\
HD 105234             & Cha-cand              & 12:07:05.5 & -78:44:28.1 & 7.17  & A9   & $101\pm 5$  & 103H   &  1.44* \\
2MJ12074597-7816064   & Cha-12,rej            & 12:07:46.0 & -78:16:06.5 & 13.11 & M3.75& \ldots      & \ldots & \ldots \\
 RXJ1207.7-7953       & Cha-32                & 12:07:48.3 & -79:52:42.0 & 12.06 & M3.5 & \ldots      & 111    & \ldots \\
 HIP 59243            & Cha-cand              & 12:09:07.8 & -78:46:53.0 & 6.56  & A6   & $96\pm 6$   &  94H   &  1.57* \\
 HD 105923            & Cha-33                & 12:11:38.1 & -71:10:36.0 & 8.31  & G8   & \ldots      & 112    &  1.99* \\
 RXJ1216.8-7753       & Cha-34                & 12:16:45.9 & -77:53:33.0 & 11.65 & M4   & \ldots      & 118    & \ldots \\
 RXJ1219.7-7403       & Cha-35                & 12:19:43.7 & -74:03:57.3 & 10.95 & M0   & \ldots      & 112    & \ldots \\
 RXJ1220.4-7407       & Cha-36; KOH 93        & 12:20:21.8 & -74:07:39.4 & 10.8  & M0   & \ldots      & 110    &  0.24  \\
 2MJ12210499-7116493  & Cha-37                & 12:21:04.9 & -71:16:49.3 & 10.21 & K7   & \ldots      & 110    & \ldots \\
 RXJ1239.4-7502       & Cha-38                & 12:39:21.3 & -75:02:39.2 & 9.21  & K3   & $103\pm 5$  & 100    & \ldots \\
 RXJ1243.1-7458       & Cha-rej; BRR 6        & 12:42:53.1 & -74:58:49.0 & 12.72 & M3.2 & \ldots      & \ldots & Triple \\
 CD-69 1055           & Cha-39                & 12:58:25.6 & -70:28:49.0 & 8.89  & K0   &  $93\pm 5$  &  99    & \ldots \\
 CM Cha               & Cha-cand              & 13:02:13.6 & -76:37:58.0 & 11.8  & K7   & \ldots      & 133    & \ldots \\
 MP Mus               & Cha-40; MP Mus        & 13:22:07.6 & -69:38:12.0 & 9.18  & K1   & $99\pm 7$   & 101    & \ldots \\
	\enddata                    
\tablenotetext{a}{Derived from Gaia DR1 parallaxes, \citep{gaia2016a,gaia2016b}}
\tablenotetext{b}{From \cite{Murphy2013}. An "H" means the parallax is from Hipparcos. Otherwise parallaxes are kinematic.}
\end{deluxetable*}

\explain{In Table 1 we added two columns. The first one contains distances for $\epsilon$ Cha members, candidates or rejected members derived from the Gaia DR1. The second column contains distances from \cite{Murphy2013}.}

\subsection{Instrument and Observing Method}

The speckle  observations reported here were obtained  \replaced{during the nights of
	January 16-18, 2016, allocated through NOAO program 15B-0268, PI CB. All 2.5 nights of this run were
	clear, with variable seeing and a slow wind}{on 2016 January
  17. The time was allocated through the NOAO  program 2015B-0268, C. Brice\~no Principal Investigator.  The sky
  was clear, with good seeing and a slow wind.}

We used the  {\it High-Resolution Camera} (HRCam) -- a fast imager designed to work at
the 4.1-m SOAR  telescope \citep{TC08}. The camera was  mounted on the
SOAR  Adaptive Optics Module  \citep[SAM;][]{tok2016a}.\footnote{\url{http://www.ctio.noao.edu/soar/content/soar-adaptive-optics-module-sam}} 
We  used the  UV  laser to
correct for turbulence,  in order to achieve a  deeper magnitude limit
and  better  resolution; this  observing  mode  was  used earlier  for
screening    {\it   Kepler-2}    variable    stars   for    companions
\citep{Kepler2}. 
The  SAM  module corrects  for  atmospheric  dispersion  and helps  to
calibrate the pixel scale and  orientation of HRCam \citep[see][]{TMH15}.  
We used mostly the $I$-band filter ($\lambda_0=788$\,nm; FWHM=132\,nm).
The
transmission  curves of  HRCam  filters are  given  in the  instrument
manual.\footnote{\url{http://www.ctio.noao.edu/\~atokovin/speckle/index.html}}

After acquiring  each target and centering  it in the  HRCam field, we
closed the  laser loop. The  overhead associated with using  the laser
guide  star (LGS)  was  only  a few  seconds  when observing  multiple
targets in the same area on the  sky. Once the LGS is centered for one
target,  no   further  adjustments   are  needed  for   the  following
targets. The  laser switches  on when the  telescope is slewed  to the
target (the software that controls laser propagation within authorized
time windows takes  care of this).  Laser interrupts  had only a minor
effect because the exposure times  were short.  The high-order AO loop
compensates for telescope  aberrations and low-altitude turbulence; it
automatically  maintains the  optimum  focus.  Residual  tip and  tilt
jitter  is  compensated in  the  data processing.  If  we  had used  a
classical CCD imager, also  available with SAM, the observations would
have been  much less efficient  because acquisition of  off-axis guide
stars would be needed for  each field.  For example, the binary survey
using SAM  with a  classical CCD  could cover only  21 targets  in one
night \citep{Tok2014}; \added{in contrast, the 47 stars of this program were observed
  in 3.2 hours.} 

Without the laser, HRCam reaches a  magnitude limit of $I \sim 12$ mag
under good seeing.  SAM provides  an increase of $\sim 1$ magnitude in
depth,  allowing  us to  go  down to  $I\sim  13$  mag targets  (Table
\ref{targets}).  We used \added{a} detector binning of 2$\times$2 \added{that
produces an} effective  pixel scale  of  30.46\,mas. For  each target,  we
acquired two cubes  of 200$^2$ binned pixels size,  covering the field
of 6\arcsec$\times$6\arcsec,  with 400 frames per  cube.  The exposure
time  was  from  0.1 to  0.2~s  per  frame,  depending on  the  target
brightness (i.e. 40 to 80~s accumulation time per data cube). Then two
more cubes  were acquired with half  the field and  a shorter exposure
(typically 0.05  to 0.1~s). These  extra narrow-field cubes  helped to
increase  the  resolution  at  the expense  of  sensitivity.   Shorter
exposures were used  for targets brighter than $I=12$  mag, which were
also  recorded without  binning. \added{Bright stars  did not  need the
  laser correction.}  Acquisition of two data cubes in each mode helps
to  confirm new  detections and  avoids artifacts  such  as occasional
cosmic rays spoiling some frames in one of the two cubes.

During \replaced{the first half of the first night, 2016 January 16, 
	and for the whole of the second night}{these observations, }  
the seeing in the  free atmosphere
reported  by  the  site  monitor  was very  good,  fluctuating  around
0\farcs3.  The total seeing  varied between 0\farcs5 and 1\arcsec. The
SAM  AO system  successfully compensated  low-altitude  turbulence and
delivered sharp images.  The median Full Width at  Half Maximum (FWHM)
of the re-centered  average images in closed loop  is 0\farcs33, while
\replaced{25}{some} data  cubes have  FWHM less  than 0\farcs25  and  80\% are
better than 0\farcs4.\deleted{On the last  night of the run, 2016 January 18,  the seeing started to
	degrade  and reached quite  large values  up to  2\arcsec. The  SAM AO
	system did not compensate such strong turbulence, so the observations
	of faint stars  could not be continued. Instead,  we observed a number
	of brighter stars in the classical speckle mode,
	without  laser. The  diffraction-limited resolution  was  reached on
	bright  stars under  poor seeing,  which degraded  only  the magnitude
	limit of the speckle technique.}

Some newly resolved pairs have been re-observed with HRCam on 2017 May
15,  this time  without  the laser and  under  mediocre seeing.   These
confirmation measurements prove  that even relatively faint companions
at $\sim$1\arcsec ~separation are  not background stars. Otherwise the
relatively large  proper motion of  40 mas~yr$^{-1}$, directed  to the
West,  would   have  changed  the  relative   companion  positions  by
0\farcs05, significantly larger than the errors.

\subsection{Data processing}

\begin{figure}[ht]
\epsscale{1.1}
\plotone{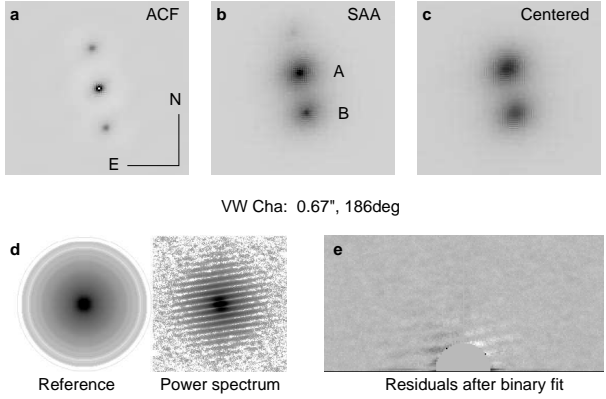}
\caption{\label{fig:example}  Example of  the speckle  data processing
  for the well-resolved binary VW~Cha.   The rop row shows the central
  fragments of  images, in  the negative scale:  the ACF (a),  the SAA
  image (b), and the centered image (c). The bottom row shows the data
  in the Fourier space: the  power spectrum and the reference spectrum
  in  the  negative logarithmic  scale  (d)  and  the residuals  after
  fitting with the  binary-star model in the linear  scale (e). As the
  power spectrum  is symmetric, only  the upper half of  the frequency
  plane is fitted; the low spatial frequencies are masked out. }
\end{figure}

The speckle  data processing described in \cite{TMH10}
was adapted to the  faint  stars  \citep[see][]{Kepler2}. 
\added{It is illustrated in
Figure~\ref{fig:example}. } As a  first  step,  power
spectra  are calculated from  the data  cubes.  While  processing each
frame,  the  bias  and   scaled  dark  signals  are  subtracted.   The
auto-correlation functions (ACFs) are computed from the power spectra.
They  are used  to detect  companions  and to  evaluate the  detection
limits.  For  each data cube,  the speckle pipeline also  delivers the
average image re-centered using optimized center-of-gravity algorithm,
and  the shift-and-add  (SAA) or  ``lucky'' image  re-centered  on the
brightest pixel in each frame,  with weight proportional to the signal
in this pixel.

Parameters of  binary and triple  stars are determined by  fitting the
power spectrum  to its  model, which  is a product  of the  binary (or
triple) star \added{power} spectrum and the reference spectrum.  We used
as a reference the  azimuthally-averaged spectrum of the target itself
in the  case of binaries wider  than 0\farcs1.  For  closer pairs, the
``synthetic''  reference  was  used   (see  TMH10).   Some  data  were
processed  using  other  observed  objects  as a  reference.   If  the
reference object  is a binary, it  is converted into a  single star by
deconvolution,  using the measured  binary parameters.   Photometry of
wide  (classically   resolved)  binaries  is   corrected  for  speckle
anisoplanatism using the centered  images.  Inspection of the centered
and SAA images helped to confirm some companions seen in the ACF.

The relatively  long exposure times   used for targets  fainter than
$I_C=13$ mag reduced the effective resolution to about 0\farcs1, while
the diffraction-limited  resolution of  0\farcs04 was reached  for the
brighter  stars.   No  appreciable  image elongation  of  instrumental
origin was  noted in  the data. Nevertheless, we  always compared  with
stars  observed  before  or  after  each  target  to  check  for  such
artifacts. The  detector orientation  and pixel scale  were accurately
calibrated   on  wide  binaries   with  well-modeled   linear  motions
\citep{TMH15}.   The  calibration parameters  are  determined with  an
accuracy of 0\fdg1 and 0.1\%, respectively.

\begin{figure}[ht]
\epsscale{1.2}
\plotone{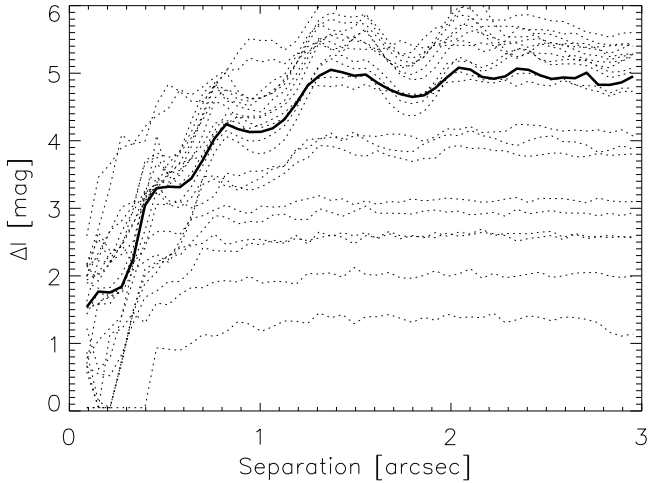}
\caption{\label{fig:det} Detection limits for 20 faint unresolved targets recorded
  with 2$\times$2 binning and 0.1\,s exposure time. The thick curve
  shows the median detection limit while the dotted lines are the
  individual limits.}
\end{figure}

The detection  limits were  estimated from the  ACFs by  computing the
variance in  annular zones and assuming that  companions brighter than
$5 \sigma$  are detectable (see TMH10).  Figure~\ref{fig:det}
shows the detection limits for  faint targets observed in closed loop.
They  vary substantially, depending  on the  target brightness  and AO
compensation quality.   The median magnitude difference  $\Delta I$ at
0\farcs15  separation is 1.75  mag, while  at 1\arcsec  ~separation it
reaches   4  mag.   Individual   $\Delta  I$   limits  at   these  two
characteristic separations  are provided  in Table~3 
for unresolved targets. Linear interpolation between these points can be used
to get the detection limits at other separations.

\begin{figure}[ht]
\epsscale{1.1}
\plotone{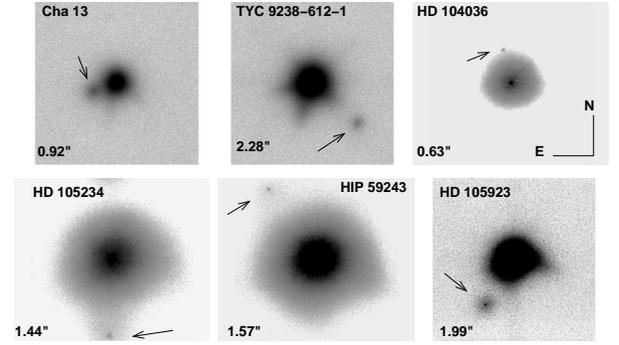}
\caption{\label{fig:wide} Images  of six newly  resolved binaries with
  faint  companions  displayed   on  arbitrary  negative  scale.   The
  separation is indicated in the  lower left corner of each image, the
  companions are marked by arrows. {\bf Cha-13 and TYC 9238-612-1} 
  are average images in
  closed AO loop, the rest are SAA images with or without AO.}
\end{figure}

\begin{figure}[ht]
\epsscale{1.1}
\plotone{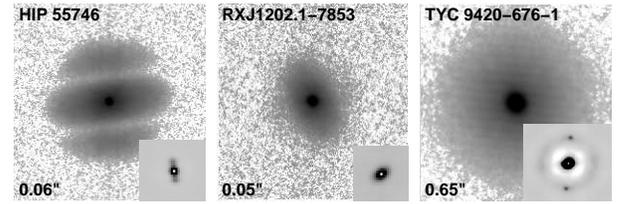}
\caption{\label{fig:close}  Power  spectra  of  three  resolved  close
  binaries displayed  on the logarithmic scale.  The  separation is
  indicated in the lower left corner; {\bf the inserts in the lower right
  corners show the central fragments of the ACFs.} }
\end{figure}

%---------------------------------------------------------
\section{Results}
\label{sec:res}

Figure~\ref{fig:wide}  shows  centered or  SAA  images  of six  newly
resolved  binaries.   Another three new  close  pairs  are illustrated  in
Figure~\ref{fig:close}  by  their power  spectra  showing fringes or elongation.

\tabletypesize{\footnotesize}
\begin{deluxetable*}{llccrrll}
	\tablewidth{0pt}
	\tablecaption{Measurements of resolved multiples in $\epsilon$ Cha \label{measurements}}
	\tablehead{\colhead{WDS} & \colhead{Name} & \colhead{Epoch}  & \colhead{Filter} & \colhead{$\theta$}   & \colhead{$\rho$} & \colhead{$\Delta m $} &  \colhead{Note} \\
	\colhead{} &	\colhead{}     & \colhead{$+2000$} & \colhead{}       & \colhead{($^\circ$)} & \colhead{($\arcsec$)} & \colhead{(mag)}  & \colhead{} 
	}
	\startdata
11080$-$7742 & VW Cha; GHE 35 AB       &       16.0474 &	I &	185.9 &		0.6670 & 	0.2 *  &     \\ 
11186$-$7936 & Cha 13                  &	16.0475 &	I &	107.8 &		0.9218 & 	3.2 *  &     \\ 
11253$-$8457 & HIP 55746                &      16.0475 &	I &	184.9 &		0.0614 & 	0.6 q  &     \\ 
            &	                        &      16.0475 &	y &	185.2 &		0.0607 & 	0.9 q  &     \\ 
            &                           &      16.9595 &        y &     187.1 &          0.0525&        0.3   &  \\
            &                           &       16.9595 &       I &     188.4 &          0.0489&        0.4   &  \\
11375$-$7648 &  RXJ1137.4-7648        &         16.0475 &       I &     5.5   &         2.8952 &        1.8   &  \\      
11415$-$7347 &	TYC 9238-612-1       &         16.0475 &        I &	226.1 &	        2.2767 & 	5.7 *  &   \\
             &                       &         17.3719 &        I &     226.3 &         2.2620 &        5.3 :  & \\      
11509$-$7411 & RXJ11509-7411; BRR 15 &  	16.0475 &	I &105.3 & 	0.9143 &	1.7 *  &     \\ 
11585$-$7749 &	HD 104036; Cha 27 &     	16.0475 &	I &10.9 &  	0.6304 &	3.6 *  &     \\ 
            &	       	           &	       16.0475 &	y &10.9 &  	0.6295 &	4.9 q  &     \\ 
            &	                   &           17.3718 &	I &10.8 &	0.6383 &	3.7 q  & \\
11585$-$7754 &RXJ115805-7754A; KOH 91 & 	16.0475 &	I &147.7 & 	0.0306 &	1.0 :  & Marginal     \\ 
11596$-$7813 & $\epsilon$~Cha; HJ 4486 Aa,Ab   &16.0475 &	y &47.6 &  	0.0592 &	0.2    & Triple    \\ 
             &                                 & 17.3717 &	y &28.8 &	0.0612 &	0.2    & \\
11596$-$7813 & $\epsilon$~Cha; HJ 4486 Aa,B   &	16.0475 &	y &239.0 & 	0.1779 &	0.1    & Triple    \\ 
             &                                & 17.3717 &	y &247.0 &	0.1731 &	0.0    & \\
12001$-$7812 &	HD 104237A; GRY 1 AF 	 &	16.0475 &	I &254.8 & 	1.3904 &	5.5 *  &     \\ 
12001$-$7811 &	HD 104237D; FGL 2 ED 	 &	16.0477 &	I &311.3 & 	4.2409 &	1.9    &     \\ 
12021$-$7853 &	RX J1202.1-7853; Cha 30 &	16.0475 &	I &111.9 & 	0.0521 &	1.2    & Tentative    \\ 
12049$-$7932 &	TYC 9420-676-1         &  	16.0476 &	I &172.3 & 	0.6565 &	3.4 *  &     \\ 
             &                       & 	       17.3718 &	I &172.3 &	0.6605 &	3.7    & \\
12070$-$7844 &	HD 105234               &	16.0476 &	I &175.0 & 	1.4408 &	4.4 *  &     \\ 
             &                         & 	17.3718 &	I &174.7 &	1.4303 &	5.6 *  & \\
12091$-$7846 &	HIP 59243               &	16.0476 &	I &31.7 &  	1.5676 &	5.6 *  &     \\ 
12116$-$7110 &	HD 105923              &	16.0476 &	I &145.5 & 	1.9893 &	4.7 *  &     \\ 
12204$-$7407 &	RXJ1220.4-7407; KOH 93   &	16.0476 &	I &9.4 &   	0.2446 &	1.8 q  &     \\ 
             &                           & 	17.3719 &	I &12.1 &	0.2448 &	1.2 :  & \\
12431$-$7458 &	RXJ1243.1-7458; BRR 6 Aa,B &	16.0477 &	I &258.1 & 	2.5246 &	2.5    &  Triple   \\ 
12431$-$7458 &	RXJ1243.1-7458; KOH 94 Aa,Ab  &	16.0477 &	I &79.8 &  	0.2379 &	1.3 q  &  Triple   \\
%            &	       &	           &	16.0477 &	I &78.7 &  	0.2357 &	1.5    &     \\  
	\enddata                    
\end{deluxetable*}

\vspace*{2cm}

	\tabletypesize{\footnotesize}
	\begin{deluxetable}{lccc}
	\tablewidth{0pt}
	\tablecaption{Detection limits for unresolved targets \label{tab:det}}
	\tablehead{\colhead{Name} & \colhead{$\rho_{min}$} & \colhead{$\Delta m(0\farcs15)$} & \colhead{$\Delta m(1\arcsec)$} \\
	\colhead{}  & \colhead{(\arcsec)} & \colhead{(mag)} & \colhead{(mag)} }
	\startdata
 HD 82879           &   0.04  &   1.6   &  5.1   \\
 CP-68  1388        &   0.04  &   2.2   &  4.8   \\
 TYC 9414-191-1     &   0.04  &   1.6   &  4.3   \\
 RXJ1123.2-7924     &   0.04  &   1.9   &  3.4   \\
% RXJ1137.4-7648     &   0.04  &  2.54   &  3.77   \\
 2MJ11432669-78...  &   0.10  &   0.9   &  2.2   \\
 RXJ1147.7-7842     &   0.04  &   1.7   &  4.0   \\
 RXJ1149.8-7850     &   0.04  &   2.1   &  4.3   \\
 RXJ1150.4-7704     &   0.04  &   1.8   &  4.6   \\
 2MJ11550485-79...  &   0.10  &   1.7   &  3.3   \\
 T Cha              &   0.04  &   2.4   &  5.1   \\
 RXJ1158.5-7754B    &   0.04  &   2.0   &  4.2   \\
 CXOUJ115908        &   0.10  &   2.2   &  2.3   \\
 HD 104237E         &   0.04  &   1.6   &  4.6   \\
 2MJ12005517-78...  &   0.04  &   1.4   &  2.6   \\
 HD 104467          &   0.04  &   0.7   &  6.1   \\
 USNOB-120144-78... &   0.10  &   2.0   &  2.9   \\
 CXOUJ120152-78...  &   0.10  &   2.0   &  3.2   \\
 RXJ1202.8-7718     &   0.04  &   2.5   &  4.5   \\
 RXJ1204.6-7731     &   0.04  &   2.3   &  4.5   \\
 2MJ12074597        &   0.10  &   2.5   &  3.6   \\
 RXJ1207.7-7953     &   0.04  &   2.6   &  3.9   \\
 RXJ1216.8-7753     &   0.04  &   2.4   &  4.6   \\
 RXJ1219.7-7403     &   0.04  &   1.8   &  4.2   \\
 2MJ12210499-71..   &   0.04  &   1.7   &  5.1   \\
 RXJ1239.4-7502     &   0.04  &   1.9   &  5.3   \\
 CD-69 1055         &   0.04  &   0.7   &  5.2   \\
 CM Cha             &   0.04  &   2.2   &  4.4   \\
 MP Mus             &   0.04  &   2.1   &  4.9   \\
	\enddata                    
	%\tablecomments{}
\end{deluxetable}

Table~\ref{measurements} lists \replaced{30 measures of 19 
	binary pairs, including 9 newly resolved ones}{31 measures of 20 
binary pairs, including 10 newly resolved ones.} 
\explain{During a closer look at the data we detected an additional new binary, RXJ1137.4-7648. 
Because it was close to the edge of the 6 arcsec FOV we had missed it during our initial inspection. 
It has now been added.}
The columns of Table~\ref{measurements} 
contain (1) the WDS-style code based on the J2000 coordinates;
(2) the star name, from Table 1 in \cite{Murphy2013}, 
(3) the Besselian epoch of observation, 
(4) the filter used,
(5) the position angle $\theta$ in degrees,
(6) the separation $\rho$ in arcseconds,
(7) the magnitude difference $\Delta m$, 
with an asterisk following if $\Delta m$ and the true quadrant
are determined from the resolved long-exposure image; a colon
indicates that the data are noisy and $\Delta m$
is likely overestimated (see TMH10 for details); 
the flag 'q' means that the quadrant is determined from the SAA image. 
In cases of multiple stars, the positions and photometry refer to the pairings
between individual stars, not the photocenters of subsystems. The last
column (8) gives short notes for some objects.

Table~\ref{tab:det}   lists  the   detection  limits   for  unresolved
targets. The  minimum separation $\rho_{\rm  min}$ in column  2 equals
the diffraction limits of 40\,mas  for all targets except the faintest
ones, where  it is  0\farcs1. The following  columns give  the maximum
detectable  magnitude difference  in the  $I$ band  at  separations of
0\farcs15 and  1\arcsec, estimated  from the ACF.  \added{Note  that the
  detection  limits in  Figure~\ref{fig:det} refer  only to  the large
  binned  data cubes,  whereas  those in  Table~\ref{tab:det} are  the
  deepest limits for each target in all observing modes.}

Overall,  we measured  \replaced{five}{eight} known  systems (including  three  triples, of
which we confirmed two) and added \replaced{nine}{ten} new binary pairs resolved here for
the first  time. The following  Subsection~\ref{sec:comments} gives comments
on  individual binaries, the Subsection~\ref{sec:echa}  is devoted  to the
multiple system $\epsilon$~Cha itself.

\subsection{Comments on  resolved systems in $\epsilon$ Cha}
\label{sec:comments}

{\it 11080-7742.} VW Cha (GHE~35~AB) \added{is a K7-M0 accreting Classical T Tauri star, 
classified as a Class II object based on its Spectral Energy Distribution \citep[SED;][]{Manoj2011}}. 
This system is resolved at 0\farcs66, without
any  trace of  the 0\farcs1  subsystem BNK~1  Ba,Bb  discovered by
\citet{Brandeker2001}  in   2000.   The  subsystem   Ba,Bb  was  also
unresolved at SOAR in 2014 and 2015, although at $\Delta J = 0.3$ mag it
should be easily detectable. It is possible that the pair Ba,Bb \replaced{closed down}{became closer}
(its  estimated period is $\sim$70\,yr).  The pair  AB moved very
little  since   its  discovery  in  1994;  its   estimated  period  is
$\sim$400\,yr.  Not a member of the association, but definitely a
young object; \added{\citet{Murphy2013} attribute it to the Cha I group.
	It is not featured in the {\it Gaia} DR1 so its true membership is still
	difficult to ascertain.}

{\it 11186-7936}. 2MASS J11183572-7935548 (Cha 13, $I=12.22$ mag, M4.5)
is   a  new   0\farcs92  binary.    \citet{Murphy2013}   suspected  RV
variability, which might mean that  it is triple, because the new wide
pair has a period of $\sim$2\,kyr. \added{In their Spitzer study of Chamaeleon,
	\cite{Manoj2011} classify this star as a Weak-lined T Tauri star based 
	on the modest H$\alpha$ emission equivalent width of 11{\AA } reported in literature
	low resolution spectra. However, the
	Spectral Energy Distribution (SED) clearly shows this is not a disk-less star,
	quite the contrary. Though it has a stellar-like SED out to $\sim 3\,\mu$m, 
	it exhibits significant excess emission at longer wavelengths, 
	with a strong $10\, \mu$m silicate emission feature, indicative of a small amount 
	of optically thin dust in an otherwise cleared gap; thus, they classify
	it as a candidate Transitional Disk (TD).
	\cite{Murphy2013} also find that H$\alpha$ is variable in Cha-13, though the line profile itself
	is not very wide, with a 10\% width of $\rm \sim 170\, km\, s^{-1}$, consistent with
	a low accretion rate of $\rm \sim 10^{-11} M_{\odot}\, yr^{-1}$.
	These characteristics closely resemble those of 
	other TDs, like CVSO-224 in Orion \citep{Espaillat2008}.
	In addition to H$\alpha$, \cite{Murphy2013} find a number of other emission lines in their
	spectra of Cha-13, like He I, [NII]$\lambda 6548/6583$, [SII]$\lambda 6716/6731$,
	among others,
	leading them to suggest the gap may have been cleared by jet/outflow activity. 
	Our discovery of a companion at $\sim 90$ AU raises the alternative interpretation that 
	the TD status is related to the binary nature of the system. At separations of
	$\lesssim 200$ AU a companion can truncate the outer disk, while
	at separations of $\lesssim$ few tens of AU the disk could be truncated
	from the inside \citep{Manoj2011}.  Indeed, studying Chamaeleon I \cite{Daemgen2016} find that there is
	a statistically significant difference in the accretor fraction between single and binary
	systems; in particular, binary systems with separations $\lesssim 100$ AU show a low $\sim 6$\% incidence
	of accretion activity. More specifically,
	a recent study of 24 TDs shows that close to 38\%
	can be explained by tidal interactions between a close binary companion and its disk,
	while the rest is likely the result of processes like disk photoevaporation, grain growth or
	planet-disk interactions \citep{RRod2016}.  
	Does the weak accretion activity and excess infrared emission in Cha-13 originate
	in one or two circumprimary disks, or maybe in a circumbinary disk? 
	Has the gap in this TD system been carved out by the binary companion? 
	These are open questions for
	this very interesting system that clearly deserves further detailed studies.}

{\it 11253-8457.} HIP~55746  is revealed as a tight  60-mas binary with
an  estimated period of  $\sim$10\,yr. This  close pair  was suspected
from  the astrometric  acceleration detected  by {\it  Hipparcos} \added{\citep{frankowski2007}}. 
The pair was  resolved again at SOAR  in 2016.96 and closed  down to about
30\,mas in  2017.34 (the elongated  power spectrum was not  fitted, no
measurement).  There is another  companion at 3\farcs54 (RST~2752), so
the  system is  triple.   The wide  physical companion,  last seen  in 1996  at
(208\degr, 3\farcs5),  is outside the field  of view, hence  it is not
detected here.  

\added{ {\it 11375-7648.}  RX  J1137.4-7648 ($I=12.2$  mag,  M2), not  a
  member of  $\epsilon$~Cha, has a  wide 2\farcs9 companion  that just
  fits in the 6\arcsec ~field. Its  image is partially truncated, so the $\Delta
  I$ is over-estimated by some unknown amount.}
  \replaced{Although \citet{Murphy2013} call it  ``equal-brightness visual binary'',
  	referencing \citet{Correia2006}, this pair is not featured neither in the
  	literature nor in the WDS.}{Although \citet{Murphy2013} call it  ``equal-brightness visual binary'',
  and it is evident as a visual pair in Digitized Sky Survey (DSS) and 2 Micron All-Sky Survey (2MASS) images,
  this pair is not featured neither in the literature nor in the WDS.}

{\it 11415-7347.}  TYC 9238-612-1 ($I=9.98$ mag, G5) has  a faint companion
at 2\farcs28. It is barely detectable at the same position in 2017.37. 

{\it 11509-7411.} RX J1150.9-7411 is a previously known binary BRR~15,
resolved here at 0\farcs91. It has not moved appreciably since
its discovery in 1994. \citet{koh2001b} found it at 0\farcs875 and
106\fdg0. 

{\it  11585-7749.}  HD  104036  ($I=6.49$  mag, A7)  has  a new  faint
companion at  0\farcs63.  The  pair is found  at the same  position in
2017.37, confirming that the  companion is physical. \added{The new companion
is  too  distant  for  explaining  the  RV  variability  suspected  by
\citet{Murphy2013}.}

{\it  11585$-$7754.} The  close binary  KOH~91 (Cha-21)  is marginally
resolved at 30\,mas. Although the resolution is tentative, this is the
only confirmation of the original discovery made by K\"ohler at
separation 0\farcs073 and magnitude difference $\Delta K = 0.75$ mag. The
system is triple considering the faint companion RXJ1158.5-7754B at
16\farcs6 distance. The estimated period of AB is 30\,kyr.

{\it  12001-7811.}   HD~104237  ($I=6.31$),  an  accreting  Herbig  Ae
spectroscopic  binary  with  a  circumbinary disk,  is  surrounded  by
several faint stars, forming a kind of mini-cluster \citep{Grady2004}.
We  resolved  the  closest   1\farcs4  pair  AF\added{\footnote{\bf  The  WDS
  component F  corresponds to the ``star 2''  in \citet{Grady2004} and
  is sometimes called ``B'',  causing confusion with the 4\farcs1 pair
  FGL~2AB listed in the WDS, which is erroneous and unphysical. 
  Based on the images presented by \cite{Grady2004} and \cite{fwg2003}, 
  the close AF pair is the same as the AB pair described by \cite{fwg2003}, 
  and there is no 4.1" A-? pair among the known components. 
  The WDS AB pair appears to be a historical artefact. 
  }}  and measured the 4\farcs24 pair ED  owing to its favorable  orientation along the
diagonal of the detector.  HD~104237 is located at 135\arcsec ~from the
$\epsilon$~Cha and is  listed in the WDS as  its companion C, although
the orbital period of  such a wide pair AB,C would be  on the order of
0.5\,Myr.    The   19-day    inner   spectroscopic   binary   resolved
interferometrically  by  \citet{Garcia2013} has  a  semimajor axis  of
2\,mas, well below the  SOAR resolution limit.  \replaced{Our observations
do not reveal  any additional close  companions to
HD~104237 itself and to two  of its satellites}{Not surprisingly,
  our observations  do not reveal  any additional close  companions to
  HD~104237 itself and to two  of its satellites because these objects
  are already well studied.}

{\it 12021$-$7853.} RX J1202.1-7853 ($I=10.5$ mag, M0) has an elongated
power     spectrum     corresponding     to    a     50-mas     binary
(Figure~\ref{fig:close}).   The resolution  is  tentative, but  likely
real  by  comparison with  other  targets  that  do not  show  similar
elongation.   This  star  was  observed by  \citet{koh2001b}  but  not
resolved, being below the diffraction limit of 0\farcs13.  \added{The RV
  variability found by \citet{Murphy2013} could be caused by the new
  close companion.}

{\it  12049$-$7932.} TYC  9420-676-1  ($I=9.7$ mag, F0)  has  a new companion  at
0\farcs65, confirmed as physical by  its repeated measurement
in 2017.37.  According to \citet{Murphy2013} it does not belong to the
association. 

{\it 12070$-$7844.} HD~105234 ($I=7.2$ mag, A9) has a new 1\farcs44
companion, confirmed as physical in 2017.37. 

{\it 12091-7846.}  HIP~59243 ($I=6.56$ mag,  A6) is resolved  at 1\farcs57.
The binary is very likely physical, but there is no second measure to
confirm this.

{\it 12116$-$7110.} HD~105923 ($J=8.3$ mag, G0) has a faint companion
at 1\farcs99. \citet{Elliott2015} also  detected this binary in 2006 at
1\farcs96 and 145\fdg1. The companion is thus physical. 

{\it  12204-7407.} RX  J1220.4-7407  ($I=10.80$ mag,  M0)  is a  known
binary KOH~93.  It was discovered in 1996 at (348\fdg4, 0\farcs296) by
\citet{koh2001b}.   It  is found  here  at  (9\fdg3, 0\farcs24).   The
estimated period is $\sim$140\,yr; the observed direct motion (21\degr
~in   30\,yr)  matches   this  crude   estimate.  \added{According  to
  \citet{Murphy2013}, the star RX J1219.7−7403 at a projected separation of
  0.14\,pc could be bound to this binary, thus making it triple system.}

{\it 12431-7459.}  RX J1243.1-7458 ($I=12.72$ mag,  M3.2), not a  member of
$\epsilon$~Cha, is resolved as a triple system consisting of the close
0\farcs23 pair  and a  fainter companion at  2\farcs5. This  triple is
already known,  designated in the WDS as  KOH~94 Aa,Ab and  BRR~6 AB.
The  inner subsystem  was discovered  in 1996  at  (85\degr, 0\farcs3)
\citep{koh2001b} and not measured since. It is found at (78\fdg3, 0\farcs23). The
orbital  motion is  slow. \added{\cite{Murphy2013} note a possible spectroscopic companion,
	which, if true, would make this a quadruple system. They
	assign this system to the more distant Cha II cloud population.}

\subsection{The triple system $\epsilon$ Cha}
\label{sec:echa}

\begin{figure}[ht]
\epsscale{1.1}
\plotone{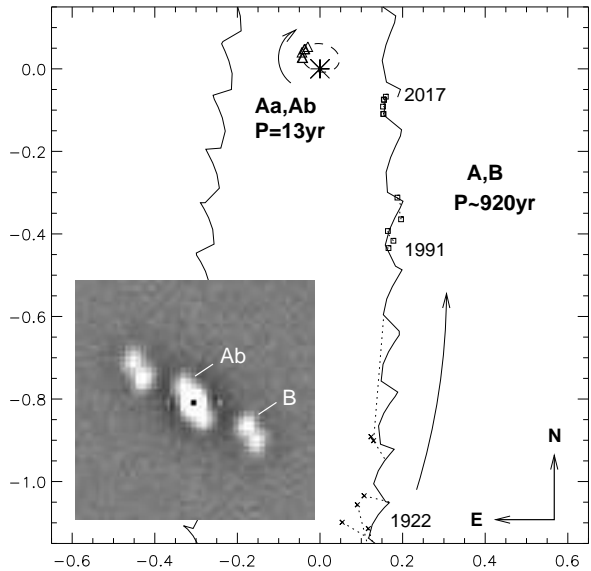}
\caption{\label{fig:EpsCha}  Tentative  orbits  of the  triple  system
  $\epsilon$~Cha with axis scale in arcseconds. Only a fragment of the
  outer  orbit is  plotted; its  wavy trajectory  reflects  the wobble
  caused by the inner subsystem. {\bf Crosses and squares connected to
    the orbit  by dotted  lines denote the  measurements of  the outer
    pair,  with some dates  indicated (the  first measurement  made in
    1835 is outside the plot).  Triangles mark the measurements of the
    inner pair.}   The insert shows the  ACF recorded at  SOAR on 2017
  May 15 {\bf where Ab and B mark the peaks  corresponding to the two
    companions.}}
\end{figure}

The  central  star  of  the  association,  $\epsilon$~Cha  (HIP  58474,
HD~104174,  B9V), is  a  known  binary, HJ~4486,  discovered  in 1835  by
\citet{Hershel1847}  at a  separation of  1\farcs6 and  position angle
179\degr. Since the discovery, the separation  of the bright  companion B
has steadily decreased to the  present 0\farcs17, with a slow increase
of the position angle to 240\degr. In 2015, the system was observed at
SOAR and unexpectedly  resolved into a tight triple  with nearly equal
components \citep{tok2016b}.

The inner  50-mas pair has turned  by 29\degr ~in two  years since its
discovery,   in   agreement   with   its  estimated   short   period.
Figure~\ref{fig:EpsCha} shows  the tentative  orbits of the  outer and
inner  pairs computed from  the available  data.  These  orbits, still
quite uncertain, are given here  only as an illustration; they are not
yet  ready for  publication.  The  provisional orbits  
match the expected masses of  these stars, about 2.5 $\rm M_{\odot}$ each.  
The short inner period means  that the inner orbit will be  constrained in a few
more years, while the millennium-long outer orbit will remain uncertain
for the lack of coverage.  Accurate speckle measurements at SOAR begin
to show the ``wobble'' in the  relative position of Aa and B caused by
the subsystem.  The amplitude of the wobble is about half of the inner
semimajor  axis  because the  components  Aa  and  Ab have  comparable
masses. Future monitoring of this interesting triple system will allow
accurate measurements  of the masses  of these young  B9V stars  and will
provide a valuable anchor point for stellar evolutionary models.

Interestingly, the  inner and outer pairs in  $\epsilon$~Cha rotate in
opposite  directions. The  provisional orbits  are  almost orthogonal,
while  the inner  orbit has  a large  eccentricity of  $\sim$0.8. This
triple system may be undergoing Lidov-Kozai cycles \added{that might
  lead to the formation of a close inner binary \citep[see the
    review by][]{Naoz2016}. }

\section{Discussion: the multiplicity fraction}
\label{sec:disc}

The     multiplicity     strongly      depends     on     the     mass
\citep{duchene_kraus2013}.   To  be  meaningful,  the  observationally
determined multiplicity fraction must  refer to the well-defined range
of primary masses, separations (or periods), and mass ratios. However,
masses and mass  ratios are notoriously difficult to  estimate for PMS
stars. As the  small sample size does not  allow accurate multiplicity
measurement  in $\epsilon$  Cha,  crude qualitative  estimates of  the
multiplicity fraction given below  seem to be appropriate.  \added{Owing
  to the  limited observational material,  we prefer not  to speculate
  about the multiplicity statistics  in $\epsilon$ Cha or compare with
  other young groups.}

We select from Table~1 thirty  members and candidate members of $\epsilon$
Cha with  spectral types of G0 or  later ($I > 8$  mag) for comparison
with  the  field  dwarfs.  There  are  6  binaries  in  the  projected
separation range  from 4 to 300  AU (1.9 dex).  All those companions
are  physical.  This  leads   to  a  raw  multiplicity  fraction  of
0.10$\pm$0.04 per  decade of  separation. As we  have not  sampled the
full  range  of  mass  ratios  owing  to the  separation-dependent
detection limit, the actual  multiplicity fraction is higher, but this
correction  depends  on the  mass  ratio  distribution  and is  highly
uncertain.   Within  errors, the  multiplicity  of  low-mass stars  in
$\epsilon$ Cha  appears to be comparable to  the multiplicity fraction
of solar-type dwarfs  in this separation range, about  0.15 per decade
\citep{duchene_kraus2013}.

\begin{figure}[ht]
\epsscale{1.1}
\plotone{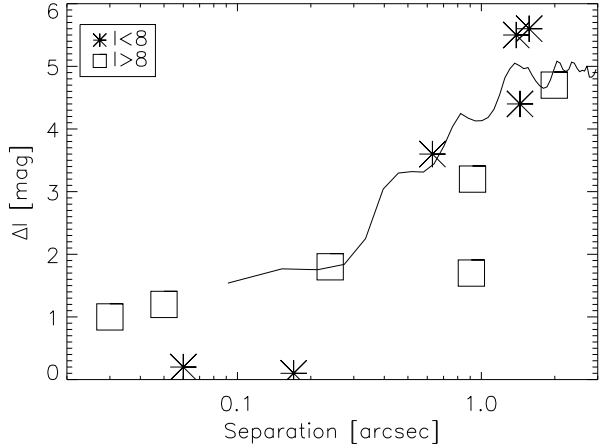}
\caption{\label{fig:bin}   Binary  companions   to   the  members   of
  $\epsilon$      Cha      association:      magnitude      difference
  vs. separation.  Squares denote stars of moderate  mass fainter than
  $I=8 $ mag, asterisks stand for five more massive and brighter members of the
  association.   The  median  detection  limit for  faint  targets  is
  plotted in full line.}
\end{figure}

We have not  detected any companions to the  four targets fainter than
$I=13$ mag.   However, the  detection limits for  faint stars  are not
very deep, while the number of  those low-mass members is too small to
make  any conclusions  regarding multiplicity  dependence  on mass.  The
faintest  resolved association  member, Cha-13,  has $I=12.2$  mag and
spectral type M4.5.

Among the five massive association members  with spectral types  A and B
($I <  8$ mag), we find  a total of  6 companions in the  the surveyed
separation range (4  to 300 AU), leading to  the multiplicity fraction
of 1.2 (0.6 per decade  of separation).  Four of those massive stars
(HD  104036,  HD  104237,  HD  105234, and  HIP  59243)  have  low-mass
companions  at separations  larger than  60 AU.   Only  $\epsilon$ Cha
itself stands out, being composed of three nearly equal B9V stars.

Figure~\ref{fig:bin}  plots  separations  of  the  binary  association
members  on the  logarithmic scale  and  compares them  to the  median
detection limit  from Figure~\ref{fig:det}. Several  faint companions
with separations of the order of  1\arcsec ~are close to the limit and
would have  been missed if  they were much closer.   Interestingly, in
Figure~3  of \citet{koh2001b}  there  are also  several binaries  with
separations between  0\farcs3 and 6\arcsec ~and  faint companions (flux
ratios less than 0.3 in the $K$  band), as well as a distinct group of
binaries with smaller separations  and roughly equal components. There
may  be a  similar pattern  in Figure~\ref{fig:bin},  where  all close
binaries have small  $\Delta I$. It would be  interesting to probe the
presence  or   absence  of  close  and  low-mass   companions  with  a
high-contrast AO, as our detection limits at small separations are not
deep enough.

We can't  help noting that  our relatively small sample  contains five
young triple systems (not  all of them are association  members).

In  summary, our work  contributes new  observational material  on the
binary statistics  in young associations and  clusters. The $\epsilon$
Cha  association appears  to be  different  in this  respect from  the
neighboring  $\eta$~Cha cluster,  where  \citet{Becker2013} noted  the
lack of  low-mass stars (a  top-heavy IMF) as  well as the  absence of
binaries with separations above 20\,AU.

\acknowledgments 

We thank the operators of SOAR D.~Maturana, P.~Ugarte, S.~Pizarro, and
J.~Espinoza for support of our program. \replaced{Detailed and constructive
  comments by the referee have been of great help to us.}{We also thank
  the anonymous referee for a detailed and constructive review that was of
  great help in improving this paper.} 

This work  used the  SIMBAD service operated  by Centre  des Donn\'ees
Stellaires  (Strasbourg, France),  and the bibliographic  references from  the
Astrophysics Data  System maintained  by SAO/NASA, and  the Washington
Double Star Catalog maintained at USNO.
This work has made use of data from the European Space Agency (ESA)
mission {\it Gaia} (\url{https://www.cosmos.esa.int/gaia}), processed by
the {\it Gaia} Data Processing and Analysis Consortium (DPAC,
\url{https://www.cosmos.esa.int/web/gaia/dpac/consortium}). Funding
for the DPAC has been provided by national institutions, in particular
the institutions participating in the {\it Gaia} Multilateral Agreement.
\added{The Digitized Sky Surveys (DSS) were produced at the Space Telescope Science Institute under U.S. Government grant NAG W-2166. The images of these surveys are based on photographic data obtained using the Oschin Schmidt Telescope on Palomar Mountain and the UK Schmidt Telescope. The plates were processed into the present compressed digital form with the permission of these institutions. }

%{\it Facilities:}  

\facility{SOAR}.

%\clearpage

%\landscape

\end{document}